\documentclass[verbatim,10pt,preprint2]{aastex}
\usepackage{graphicx}
\DeclareGraphicsExtensions{.pdf,.png,.jpg}

\gdef\sb{mag\,arcsec$^{-2}$}
\lefthead{Abraham \& van Dokkum}
\righthead{Ultra Low Surface Brightness Imaging with the Dragonfly Telephoto Array}
\slugcomment{Submitted to Publications of the Astronomical Society of the Pacific}
\begin{document}

\title{Ultra Low Surface Brightness Imaging with the Dragonfly Telephoto Array}

\author{Roberto G. Abraham\altaffilmark{1} \& Pieter G.\ van Dokkum\altaffilmark{2}}
 
\altaffiltext{1}
{Department of Astronomy and Astrophysics, University of Toronto,
50 St. George Street, Toronto, ON, Canada M5S~3H4}
\altaffiltext{2}
{Department of Astronomy, Yale University, 260 Whitney Avenue, New Haven, CT, USA 06511}

\begin{abstract}
We describe the Dragonfly Telephoto Array, a robotic imaging system optimized for the
detection of extended ultra low surface brightness structures. 
The array consists of eight Canon 400\,mm $f/2.8$~L~IS~II~USM telephoto lenses coupled to
eight science-grade commercial CCD cameras. The lenses are mounted on a common framework
and are co-aligned to image simultaneously the same position on the sky. The system
provides an imaging capability equivalent to a 0.4\,m aperture $f/1.0$ refractor with a
$2.6^{\circ}\times 1.9^{\circ}$ field of view. The system is driven by custom software for
instrument control and robotic operation. Data is collected with non-common optical paths
through each lens, and with careful tracking of sky variations in order to minimize
systematic errors that limit the accuracy of background estimation and flat-fielding. The
system has no obstructions in the light path, optimized baffling, and internal optical
surfaces coated with a new generation of anti-reflection coatings based on sub-wavelength
nanostructures.  As a result, the array's point spread function has a factor of $\sim 10$
less scattered light at large radii than well-baffled reflecting telescopes. As a result,
the Dragonfly Telephoto Array is capable of imaging extended structures to surface
brightness levels below $\mu_B = 30$\,mag\,arcsec$^{-2}$ in $\sim$10h integrations
(without binning or foreground star removal). This is considerably deeper than the surface
brightness limit of any existing wide-field telescope. At present no systematic errors
limiting the usefulness of much longer integration times has been identified. With
longer integrations (50-100h), foreground star removal and modest binning the Dragonfly
Telephoto Array is capable of probing structures with surface brightnesses below $\mu_B =
32$ \,mag\,arcsec$^{-2}$. 
The detection of structures at these surface brightness levels may hold the key to solving
the `missing substructure' and `missing satellite' problems of conventional hierarchical
galaxy formation models. The Dragonfly Telephoto Array is therefore executing a
fully-automated multi-year imaging survey of a complete sample of nearby galaxies in order
to undertake the first census of ultra-faint substructures in the nearby Universe.

\end{abstract}


\section{Introduction}

In this article we describe a telescopic imaging system that has been developed in order
to explore the Universe at surface brightness levels below $\mu_B = 30$
\,mag\,arcsec$^{-2}$. Our primary goal is to test predictions that at very low surface
brightness levels galaxies display a wealth of structures that are not seen at higher
surface brightness levels.

In a dark energy-dominated cold dark matter ($\Lambda{\rm CDM}$) Universe with largely
scale-invariant dark matter halos, $L\gtrsim L_*$ galaxies are thought to be very
efficient in accreting smaller neighbors.  This infall and accretion process leaves
tell-tale features: streams left behind in the wake of the accreted satellites, and tidal
distortions and tails. These features should have lifetimes of up to half the Hubble time,
so every large galaxy in the Universe is expected to live in an extended, irregular tidal
debris field
\citep[e.g.,][]{dubinski:96,moore:99,naab:07,kazantzidis:08,cooper:10,cooper:13}.

Although tidal features in galaxies have been identified and studied for many decades
\citep[e.g.,][]{toomre:72}, a minority of elliptical galaxies and the vast majority of
spiral galaxies appear stubbornly unperturbed, even in deep images \citep[e.g.,][and many
other studies of individual objects or small
samples]{barton:97,fry:99,dokkum:05,atkinson:13}. A possible explanation for this `missing
substructure problem' is that we have not yet probed the required surface brightness
limits to detect the predicted structures. Detailed simulations of tidal debris around
Milky Way-like galaxies \citep{johnston:08,cooper:10} and elliptical galaxies
\citep{naab:07} indicate that the majority of accreted stars live at surface brightness
levels $\gtrsim 29$\,\sb. However, deep imaging surveys with standard reflecting
telescopes typically reach surface brightness levels of `only' $\sim 28$\,\sb\
\citep[e.g.][]{dokkum:05,tal:09,martinez:10,atkinson:13}. Therefore it is natural to
suspect that the `missing substructure problem' finds its origin in the limitations of our
telescopes for undertaking low surface brightness imaging. A related issue, the `missing
satellites problem' (in which the number of dwarf galaxies in the local group is orders of
magnitude lower than expected from numerical simulations) may have a similar solution, if
the missing satellites have very low surface brightnesses.

Imaging to surface brightnesses fainter than $\sim28$\,\sb\ has proven very difficult. It
is rather striking to note that the present limit is not significantly deeper than that
reached by researchers undertaking very long photographic integrations four decades ago
\citep[e.g.][]{kormendy:74}. Over the same period of time, revolutionary advances in
detector technology have pushed back the ground-based limit for galaxy detection by over
five magnitudes, transforming the study of intrinsically faint but relatively small targets.

The reason for the slow progress in lowering the limit for the detection of low surface
brightness structures is due to the fact that the limit is not defined by photon
statistics, but rather by systematic errors. The most obvious of these is the error
introduced by imperfect flat fielding. Flat fielding errors need to be smaller than
0.1\,\% to reach $\sim 30$\,\sb. This is challenging, but it is achievable if careful
procedures are adopted. More difficult problems that must be overcome include ghosting and
diffraction, which together result in a complex and spatially variable low-surface
brightness point spread function (PSF) at large angles. This faint component of the PSF is
ignorable in most applications, but it plays a central role at low surface brightness
levels. For example, \citet[][]{slater:09} show that even with a telescope that is highly
optimized for low surface brightness imaging, diffraction from the secondary mirror
assembly, coupled with scattering and internal reflections in the optical train and
micro-roughness in aluminized optical surfaces, produces a complex emission structure that
fills the entire imaging frame below $\sim 29$\,\sb. Interestingly, it appears that
several of the the basic design trades that make large telescopes possible (in particular,
obstructed pupils and reflective surfaces) define the fundamental systematic errors that
make pushing to very low surface brightnesses so difficult.

At present the surface brightness levels needed to test the missing substructure problem
have only been approached on a few occasions. Mihos and collaborators have imaged the
center of the Virgo cluster down to $\sim 29$\,\sb\ using extremely long exposures with
the low surface-brightness-optimized, 0.9\,m Burrell Schmidt telescope on Kitt Peak. These
authors discovered a wide array of tidal features and complex structures in the
intra-cluster light \citep{mihos:05}. The same group has probed M101 down to comparable
limits using the same telescope and ray-traced removal of contaminating stellar halos
\citep{mihos:13}. Even fainter limits have been reached in the Local Group, through counts
of individual stars. Both the Milky Way and M31 are embedded in a complex network of
streams and tails, which extend out to several hundred kpc
\citep{ibatam31:01,belokurov:07,mcconnachie:09}. The imaging system described in this
paper should be capable of detecting and characterizing such structures --- if they exist
--- around any galaxy in the nearby Universe.

\section{Concept}
\label{concept.sec}

The optics of a low surface brightness-optimized telescope should have the following
characteristics: no reflective surfaces, because dust and micro-roughness on metallic
coatings backscatter light into the optical path\footnote{This is one reason why many
coronagraphs and some high-contrast imaging systems eschew reflective optics; see
\citet{nelson:08}.}; an unobstructed pupil, because any central obstruction causes
diffraction which moves energy into the wings of the PSF; nearly perfect anti-reflection
coatings, so that ghosts and flaring do not strongly pollute the focal plane; and a small
(fast) focal ratio, as the imaging speed for extended structures much larger than the
resolution limit depends on the focal ratio, not the aperture.\footnote{A standard 50\,mm
f/1.8 consumer camera lens has the same `light gathering power' as the  Keck telescope
when expressed as the number of photons per unit detector surface. The difference is that
Keck images an area of the sky that is $40000\times$ smaller. We go into this in greater
detail in Section 5.}

Combined, these characteristics describe a fast {\em refracting} telescope. Large
refractors are not commonly used in contemporary astronomical research, and the largest
remains the 1.02m $f/19$ Yerkes achromat. The textbook limitation on large refractor
telescopes is chromatic aberration, but this aberration can be well controlled using an
appropriate combination of modern glasses (and glass-like crystals), and other issues
provide the main challenges in building large refractors. These include the difficulty of
supporting large and thin optics at their edges, and the requirement that the transparent
materials used be internally very homogeneous. Overcoming these limitations is expensive
at large apertures. As a result, at present the main place for large and fast refractive
optics in astronomical research is within the re-imaging cameras of large reflecting
telescopes. Several of these cameras have multiple lenses bigger than the objective lens
of the Yerkes telescope.

Nevertheless, very fast refracting telescopes do exist in large numbers, and they are
called telephoto lenses. The fastest commercial telephoto lenses have focal ratios of
$f/2.8$ and focal lengths of up to 500\,mm. These fast, long lenses have been developed
for professional sports and wildlife photographers, who often need to freeze distant
action in low or mixed lighting conditions. The best lenses have very low optical
distortions, and because they must sometimes be used when pointed near the sun, many are
extremely well baffled. Invariably they feature a fully enclosed structure with a large
hood which also assists in controlling scattered light.

A potential downside to these lenses is that they have many optical elements, which makes
it difficult to control internal reflections. However, the latest generation of Canon
lenses features the first commercialized availability of nano-fabricated coatings with
sub-wavelength structure on optical glasses. The surface of these coatings resembles a
grid of cones whose separation is less than a wavelength of light. An incoming wavefront
averages over an increasing impact factor of coating material as it propagates through to
the glass, yielding an effective continuous variation in the index of refraction and over
a factor of ten improvement in the suppression of scattered light and internal
reflections, even on steeply curved optical surfaces. Remarkably, as a result of these
coatings, in terms of overall optical quality over a very wide field, at present no
optically fast astronomical telescope is equal to the latest generation of commercial
telephoto lenses that use this technology.

Commercially available telephoto lenses are all of relatively small aperture. As we have
already noted, imaging speed to a given surface brightness limit depends on focal ratio
and not on the aperture. Aperture is certainly very significant, however, since it defines
the focal length at a given focal ratio, and the focal length must be adequate to allow
the desired angular resolution to be achieved. The needed angular resolution needs to be
traded off against the desired field of view. A field of view of several degrees is
desirable, because nearby galaxies are enormous at low surface brightness levels. For
example, extrapolation of the profile of typical Virgo cluster elliptical suggests angular
diameters of over a degree at surface brightness levels around 30\,\sb. Ideally, such
large areal coverage would be achieved without the need for mosaiced detectors, which can
add significant challenges to low surface brightness imaging. 

\begin{figure}[htbp]
\epsscale{1.05}
\vspace{-0.3cm}
\plotone{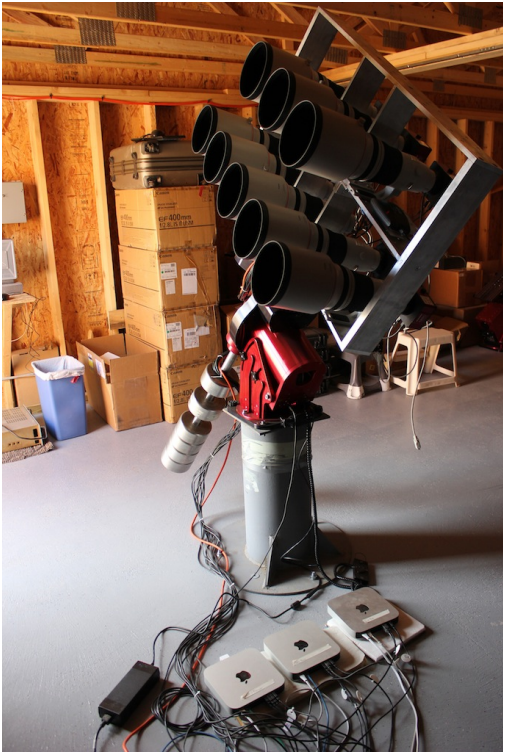}
\vspace{-0.4cm}
\caption{\small
The Dragonfly Telephoto Array installed at the New Mexico Skies telescope hosting
facility. The array consists of eight Canon 400/2.8~L~IS~II~USM telephoto lenses attached
to eight SBIG STF-8300M CCD cameras, all carried by a Paramount ME-II german equatorial
mount. The system is controlled by the three computers shown in the foreground.
\label{telescope.plot}}
\end{figure}

The Dragonfly Telephoto Array\footnote{We chose to name the concept the Dragonfly
Telephoto Array for the following reasons. (1) We took inspiration from the the compound
eyes of insects when developing the central ideas behind the array; (2) The operational
concepts that form the basis of the critical sub-wavelength nanostructure coatings were
discovered by researchers investigating the unusually high transmittance of insect wings;
(c) One of the authors really likes taking pictures of Dragonflies.} concept (shown in
Figure~\ref{telescope.plot}) is an attempt to balance these factors. It is a wide-field
visible-wavelength imaging system with significant aperture created by synthesizing a
``compound eye'' with multiple Canon 400 mm $f/2.8$ IS II telephoto lenses. This
particular lens model is the the longest commercially-available $f/2.8$ lens featuring the nano-fabricated
coatings. Each lens has an aperture of 143mm.\footnote{The only telephoto lens we are
aware of with a larger aperture is the Sigma $f/2.8$ 200\,mm -- 500\,mm zoom lens,
nicknamed ``Bigma''.} By creating an array, the effective aperture is increased while the
focal length remains fixed at 400\,mm, thus further increasing the imaging speed of the
system. Each lens is outfitted with its own monolithic CCD camera, and by combining the
light from all the cameras the effective aperture for $N$ lenses is $\sqrt{N}\times
143$\,mm and the effective focal ratio is $2.8/\sqrt{N}$. While the diffraction-limited
angular resolution of the array remains that of a single 143mm aperture lens, this is
small enough ($\sim0.65$\arcsec\, at 450nm) to be seeing-limited most of the time at most
dark sites in the continental USA.

\section{The Dragonfly Telephoto Array}

We have built and are operating the Dragonfly Telephoto Array with $N=8$ lenses as a
collaboration between the University of Toronto and Yale University. In this section we
will describe the array in its present 8-lens configuration. The present system can
accommodate up to 15 lenses with only minor modifications, and we expect to increase $N$
gradually with time. Even larger systems can be constructed by upgrading the existing
mount.

\subsection{Lens and Camera Components}

Canon 400\,mm f/2.8 IS II lenses form the heart of the Dragonfly Telephoto Array. The
first generation of these lenses came out in 1996. The second generation features the
nano-fabricated coatings discussed in \S\,\ref{concept.sec}, and became available in
August 2011. The optical design of the lenses is proprietary, but published sources
indicate that the latest model has 17 optical elements arranged in 13 groups. Each lens
has an internal autofocusing motor (which our system takes advantage of) and an image
stabilization (``IS'') unit to compensate for small motions in hand-held operation (which
our system presently does not use -- image stabilization is permanently switched off in
our application). Each lens can be stopped down with a diaphragm, enabling operation from
$f/2.8$ to $f/32$. We operate the lenses at the widest aperture, $f/2.8$, only. At $f/2.8$
the aperture of each lens is 143 mm. The relative angular positions of lenses are slightly
offset from each other in order to facilitate the data reduction process (since residual
ghosting and flaring is displaced differently on each lens).

\begin{figure}[htbp]
\epsscale{1.0}
\plotone{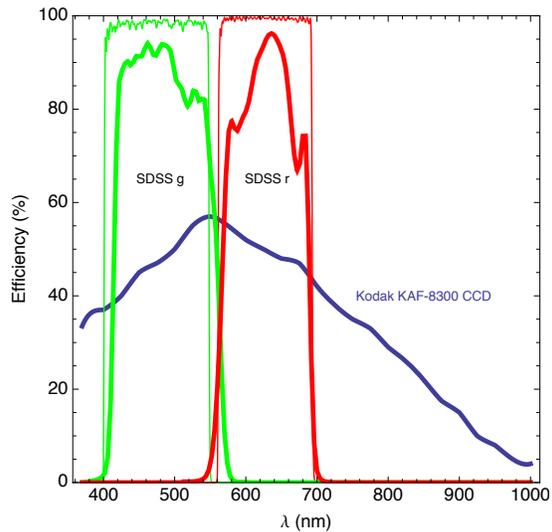}
\caption{\small 
Sensitivity of the Kodak KAF-8300 CCDs with the transmission curves of the $g$ and $r$
filters overplotted. These filters are inserted into the drop-in filter holder slots of
the Canon lenses. Green and red curves drawn with thin lines are manufacturer-supplied 
laboratory measurements of throughput for our Astrodon $g$ and $r$ filters. These
were determined using
a collimated beam. Green and red curves drawn with thick lines
are laboratory measurements for the CFHT MegaCam SDSS filters made with an $f/8$ converging beam.
\label{sens.plot}}
\end{figure}

Each of the eight lenses in the array is connected to a Santa Barbara Imaging Group (SBIG)
STF-8300M CCD camera. These cameras have Kodak KAF-8300 CCD detectors, which have a $3326
\times 2504$ pixel format. The quantum efficiency of the cameras is shown in Fig.\
\ref{sens.plot}, and peaks at 58\% at 550nm. The mean efficiency of the detector + filter
combination is 34\% in g-band and 43\% in r-band. The pixel size is $5.4\,\mu$m, which
corresponds to $2\farcs 8$/pixel. With 400mm lenses the field of view of the system is
$2.6^{\circ} \times 1.9^{\circ}$. The detectors are cosmetically excellent and have no bad
columns. Illumination of the field is quite uniform, dropping by only $\sim20\%$ from
centre to corner. The read noise of each detector is around 10 electrons (varying by
$<10$\% from detector to detector) and the gain is preset by the manufacturer to 0.37
electron/ADU. The cameras are temperature regulated using a thermo-electric cooler.
Operating at $-10^{\circ}$\,C the measured dark current is around 0.04 electron/s (varying
negligibly from detector to detector). Dark noise is removed using dark frames obtained
roughly every 90 minutes, although in practice the dark frames are very stable from
night-to-night.

Each lens is equipped with a filter via an internal drop-in filter holder. We have
obtained custom SDSS $g$ and $r$ filters from Astrodon Imaging, Inc. The filter curves are
overplotted in Fig.\ \ref{sens.plot}. We simulated the signal-to-noise (S/N) ratio in
filtered and unfiltered images using the quantum efficiency curve and template spectra for
the sky emission, sky absorption, and an early-type galaxy. We find that the S/N ratio is
approximately equal in $g$ and $r$ in equal exposure times. Compared to an unfiltered
image the S/N is approximately $\sqrt{2}$ times lower in each filter; that is, not much
could be gained by using broader filters than $g$ and $r$. In typical use, half the lenses
image in $g$ and half in $r$. This minimizes the impact of systematic errors when
determining colours, since the colours are obtained using data taken contemporaneously in
both filters. Typical integration times are 600s in both $g$ and $r$, which assures that
all exposures are sky-noise limited.

Because of the speed of the lenses, maintaining precise focus is a challenge. At $f/2.8$
the linear diameter of the airy disk on the detector is $d=3.1\mu$m at 450nm. The depth of
focus is $2 f d$, corresponding to 17$\mu$m. The opto-mechanical design of the lenses does
not incorporate temperature compensation at this level, and focus changes become
noticeable when the ambient temperature varies by as little as $1^\circ$C. Therefore,
frequent and robust focusing is essential. The Dragonfly lenses are focused using their
internal autofocusing motors. These motors are driven by custom adapters, made by Birger
Engineering Inc, which connect the SBIG cameras to the lenses. The adapters have no optics
and do double duty as spacers necessary to project the focal plane onto the CCD. The
internal stepping motors are fast and allow movements of 20,000 digital set point
increments (covering the full mechanical motion of the internal focus mechanism) in under
1s. A digital step corresponds to about 3.15$\mu$m of focus motion. Focus is determined
automatically using scripts which take a set of very short integrations in each camera and
run each resulting frame through SExtractor \citep{bertin:96}. The scripts fit a parabola
to the run of full-width at half-maximum (FWHM) vs. digital set point values and leave the
focus set at the minimum FWHM position. The initial focus positions for the eight lenses
are determined in a 24 point focus run (taking about seven minutes) during twilight at the
start of each night. The focus positions are then adjusted with short focus runs every 90
minutes (each taking under 5 minutes). Temperature is monitored throughout the night and
in between focus runs the focus is adjusted after each integration in an open loop fashion
using a temperature model we have built up using archival focus positions\footnote{ It is
interesting to note that although each lens operates at a focal ratio $f/2.8$ the total
system operates at an effective focal ratio of $f/1.0$. A monolithic camera operating at
$f/1.0$ would have a depth of focus of only 2$\mu$m, and maintenance of of accurate focus
would be much more challenging than it is with the Dragonfly Telephoto Array.}.

\subsection{Structure and Mounting}

In terms of pointing, slewing, and other mount control operations the Dragonfly Telephoto
Array is effectively a single telescope. The lenses and cameras are mounted together on a
rigid boxlike aluminium framework. The main structural elements of the framework are
constructed from 25mm thick aluminium struts. A finite element analysis was used to model
this framework prior to deployment in order to ensure that the main bending mode of the
framework is predominantly orthogonal to the optical axes of the lenses when the structure
is pointed at targets at low airmass, so that any residual flexure can be guided out by a
single auto guider sharing the framework with the lenses. In practice, at low airmass the
main sources of flexure are the stock mounting feet used to mount the lenses onto the
framework. Use of these feet alone resulted in significant flexure, ruining around 30\% of
the integrations. This was fixed by augmenting the stock mounting feet with three-point
ring harnesses that grip the lenses firmly near their front elements. After this minor
change flexure is negligible in 10 minute integrations taken within 45 degrees of the
zenith\footnote{At lower air masses flexure is noticeable in 10 min integrations and
integration times are generally kept to less than 5 minutes. For our galaxy survey there
are always ample targets at low airmass, but in future we will improve the stiffness of
the structure to better enable side-projects to be undertaken with the Dragonfly Array.}.

The lens framework is mounted on a Paramount ME II german equatorial mount manufactured by
Software Bisque, Inc. The total weight of the system (mounting framework + lenses + cameras +
cabling) is $\sim 80$kg, which is easily handled by this equatorial mount, which has a conservatively
rated instrument capacity of 110kg. When used with short focal length instruments (such as
telephoto lenses) the Paramount ME II  is likely to be able to support an instrument load
of over 140kg while maintaining excellent tracking accuracy. In addition to its rated
capacity, the main feature of this mount that makes is suitable for robotic operation is
its incorporation of an absolute homing sensor on each axis, which means it can recover
its absolute position on the sky after a power failure or other untoward incident.

All-sky pointing of the equatorial mount is accurate to better than 20 arcsec after the
incorporation of a 300 position pointing model constructed with the {\tt TPOINT} mount
modelling software \citep{wallace:94}. In order to refine pointing even further, our
control software does an astrometric plate solve of a short integration after each long
slew, resulting in absolute positions accurate to a few arcseconds. After an
empirically-derived periodic error correction the root-mean-square tracking error of the
mount is below 1.5 arcsec. Because our pixels are 2.85 arcsec in size, stars are generally
quite round even in unguided 600s integrations. However in order to obtain the best
possible image quality an autoguider is generally used. The latest version of our control
software automatically finds suitable guide stars and our reduction pipeline rejects
frames in which the axial ratio of stars deviates from perfect roundness by more than 15\%
over most of the frame.

\subsection{Functional Operation and Control Software}

\begin{figure*}[htbp]
\epsscale{2.0}
\vspace{-0.4cm}
\plotone{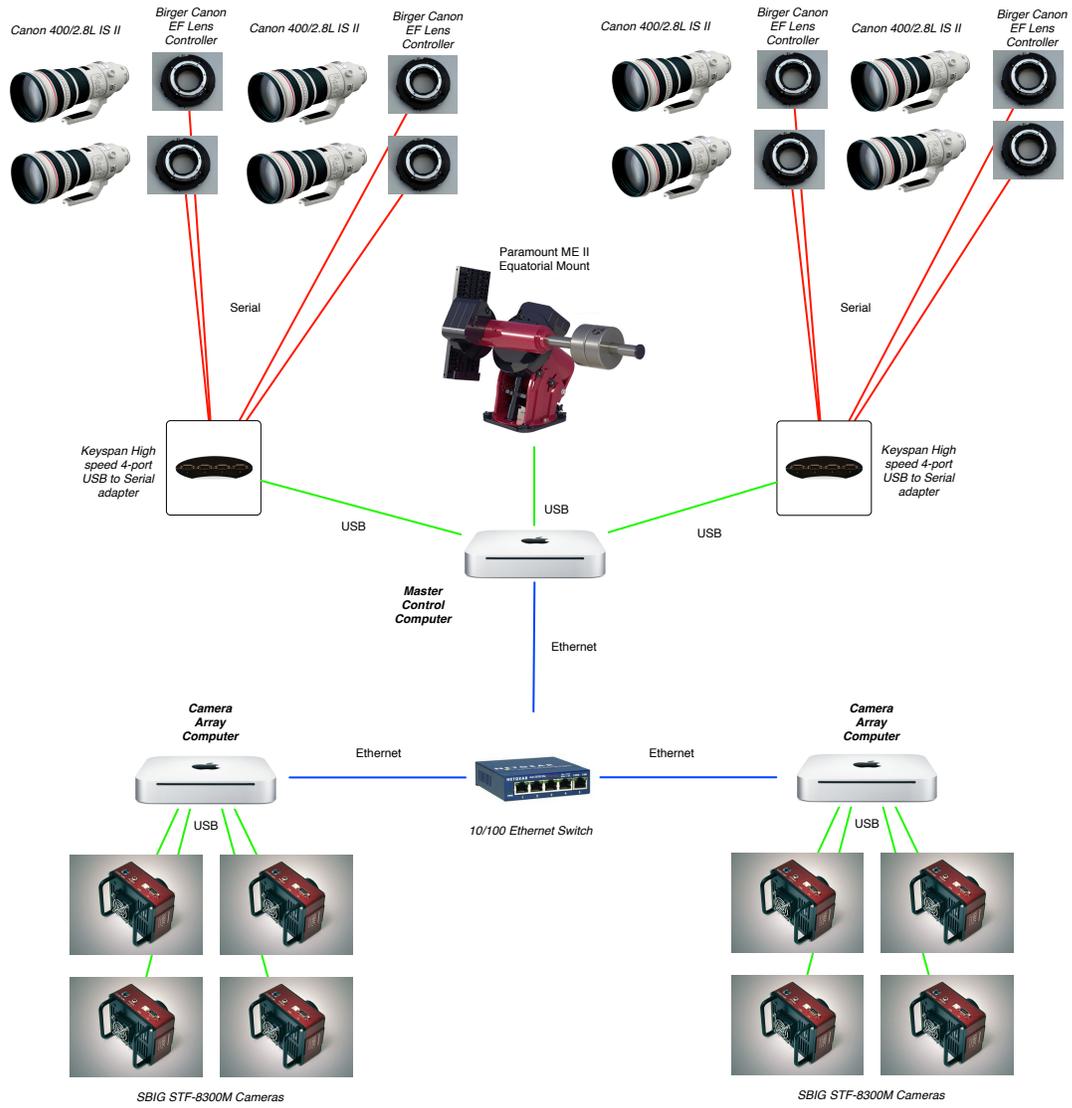}
\caption{\small
A functional overview of the Dragonfly Network Array. Serial connections are shown in red,
USB connections are shown in green, and ethernet connections are shown in blue. The lenses
are mechanically connected to the CCD cameras via focus controllers which in turn are
connected to the master control computer via serial lines. Each group of four cameras is
connected to a separate camera control computer via USB connections. Interprocess
communication between all computers occurs via TCP/IP sockets.   See text for details.
\label{network.plot}}
\end{figure*}

A functional overview of the Dragonfly Telephoto Array is presented in Figure\
\ref{network.plot}. The system is controlled by three separate computers which communicate
with each other over TCP ports using software we have written and released into the public
domain\footnote{https://github.com/robertoabraham/ProjectDragonfly}. In practice, the user
is shielded from most of the details shown in the Figure.

The user of the Dragonfly Telephoto Array array interacts only with a single computer
known as the {\em master control computer}. This is a UNIX-based system running the Mac OS
X operating system. The only tasks needed to be undertaken by the user are to log in to
this computer in the afternoon, edit and execute a small shell script (indicating the
desired targets for the night's observing) and then download the data the following
morning. A sample shell script is shown in Appendix A, which documents the required
procedure. All other activity undertaken by the Dragonfly Telephoto Array is fully
automated. Behind the scenes, the master control computer interfaces with a weather
monitor to determine whether it is worthwhile acquiring data, operates the focuser
controllers to focus the array, sends pointing and tracking commands to the equatorial
mount and autoguider (via {\tt TheSkyX}, a proprietary program written by the mount
manufacturer) and sends commands to separate computers over the network to control the
operation of the cameras. Each computer controls up to four cameras\footnote{The limitation of
at most four USB cameras per computer is set by the SBIG Universal Library, which provides
the camera control API against which our software is written.}.
During the night, images from the observatory's all-sky camera,
the results from automated focus runs, and graphical summaries of image quality and other
information are emailed at regular intervals to a special account. These constitute a
comprehensive night log.

Each camera control
computer runs a high-level network-aware server which listens for commands from the master
control computer and returns diagnostic information to it. The four cameras on each camera
control computer are read out serially (taking $<30$s to readout all four cameras), but
each camera control computer operates in parallel with the others. Therefore an arbitrary
number of additional lenses and cameras can be added to the array without adding any extra
overhead. The only constraint on adding additional lenses and cameras to the system is
that an additional camera control computer must be added for every four cameras added.

\subsection{Site}

The Dragonfly Telephoto Array is located at the New Mexico Skies telescope hosting
facility\footnote{http://www.nmskies.com/} near Cloudcroft, NM. The site is in the
Sacramento Mountains at an altitude of 2200 m. It is located about 30km from Apache Point
Observatory and is significantly darker, as it is further away from the town of
Alamogordo. Although no long-term averaged measurements of the sky brightness are
available, the site appears to be one of the darkest in the Continental
USA.\footnote{http://www.jshine.net/astronomy/dark\_sky/} The seeing at the New Mexico
Skies site is rarely sub-arcsecond (and typical long-exposure seeing is 1.5 -- 2 arcsec).
Our images are under sampled and the site seeing is not a significant limiting factor in
our observations.

New Mexico Skies is host to a large number of small robotic telescopes that are mostly
owned and operated by advanced amateur astronomers, although other institutional clients
include Caltech, NASA, and NOAO.  The Dragonfly Telephoto Array is presently mounted on a
steel pier in a large roll-off roof shed which is shared by about ten other telescopes.
The roof is closed during the day and in adverse weather conditions. The facility is
responsible for opening and closing the roof and for basic maintenance of the Dragonfly
Telephoto Array, including regular nonabrasive cleaning of optical surfaces using a ${\rm
CO}_2$ snow cleaning system.

\section{Image Quality}

\subsection{Resolution and field uniformity}

\begin{figure*}[htbp]
\epsscale{2.0}
\plotone{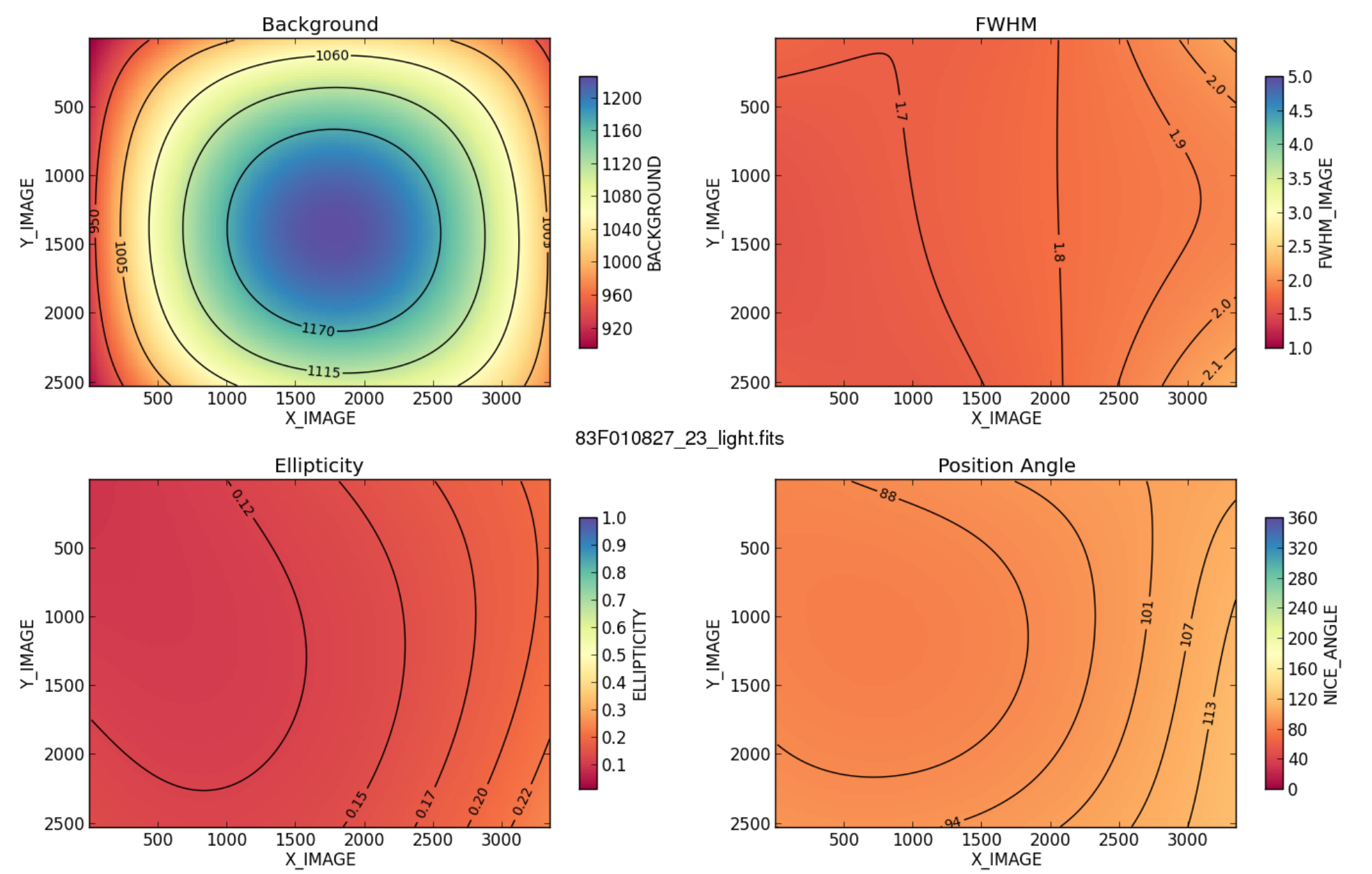}
\caption{\small
Image quality maps obtained from a typical  raw data frame, illustrating the uniformity
over the field from a single lens-camera unit.  The distribution of quantities shown were
obtained by fitting a 3rd order polynomial to results obtained by SExtractor
\citep{bertin:96} for $\sim500$ unsaturated stars. The target  was a low galactic latitude
field imaged for 600s. (Top left:) Background in counts. The illumination drops by
approximately 20\% from centre to edge. (Top right:) Full-width at half maximum (FHWM) in
pixels. At 2.85 arcsec/pixel the images are under sampled. FWHM values were determined
using SExtractor's {\tt FWHM\_IMAGE} parameter. Stars perfectly centered on a pixel have
${\rm FWHM} \sim 1.5$ pixel (the lowest possible using SExtractor's algorithm for
estimating FWHM in very undersampled images). The average FWHM is below 2 pixels over the
entire frame. Note the mild gradient in image quality. This is due to a combination of
intrinsic aberrations in the optical design of the lens, a modest amount of field
curvature and slight tilt in this particular camera-lens unit's focal plane. The maximum
FWHM degradation due to the tilt is 0.4 pixel across the frame and focal plane is tilted
at an angle of about 90 degrees. (Bottom left): ellipticity from stellar sources. The
ellipticity is below 15\% over most of the field. (Bottom right): Position angle in
degrees relative to the X-axis. At the orientation of the camera the prevailing elongation
is along the RA axis, indicative of slight tracking imperfections that were not fully
guided out. 
\label{quality.plot}}
\end{figure*}

An analysis of the image quality delivered by a typical lens-camera unit is presented in
Figure~\ref{quality.plot}. This analysis is based on a raw data frame, and shows most of
the salient features (including common imperfections) in the data. The figure displays
four panels corresponding to the background sky level, full-width at half maximum of the
point spread function (PSF), ellipticity of the PSF, and position angle of the PSF.
Distributions of these quantities were obtained by fitting 3rd order polynomials to
results obtained by SExtractor \citep{bertin:96} for $\sim500$ unsaturated stars in the
image. The FWHM values were estimated using the {\tt FWHM\_IMAGE} parameter returned by
SExtractor after filtering out galaxies and saturated stars using the {\tt CLASS\_STAR}
and {\tt FLAGS} parameters. Stellar images perfectly centred on a pixel have a ${\rm FWHM}
\sim 1.5$ pixel, as expected for under-sampled images of $\sim2$ arcsec FWHM stars imaged
by an optical system whose intrinsic angular resolution is higher than the 2.85 arcsec
sampling resolution of the detector.

The illumination pattern is well-centered on the chip, indicating good overall collimation
of the optical components in the lens. The illumination drops by approximately 20\% from
centre to edge. There is a small gradient in the delivered FWHM values in the top-right
panel. Variations in FWHM at the 10\% level are commonly seen, and they usually find their
origin in the mechanical precision of the interface between the lenses and the cameras. At
$f/2.8$ even slight mis-collimation leads to visible image degradation, and this is
exacerbated by the fact (already noted) that the depth of focus is rather shallow in fast
optical systems. Precise perpendicularity between the CCDs and the optical axes of the
lenses is required for the best image quality. In the example shown a slight tilt in the
focal plane is seen. This only impacts the image quality at a low level and it has
negligible impact on our science goals\footnote{One of the authors is congenitally unable
to leave well enough alone, so at some point in the future we expect to shim the focus
controller's mechanical spacers until the tilt in the focal plane contributes $<0.1$ pixel
to the total FWHM budget.}. Another minor source of image degradation is imperfect
tracking or guiding, which manifests itself by slightly out of round stars with position
angles oriented in the Right Ascension direction. Tracking is sufficiently good that
autoguiding is not always required, but for the sake of caution we generally do autoguide,
although with quite long integration times (5--10s) which ensure a guide star is always
available. Stellar images on our raw data frames are typically quite round, with
deviations at the 10--15\% level, which is adequate for our science. In future we will
probably upgrade the autoguider to improve image quality further.

\begin{figure}[htbp]
\epsscale{1.0}
\vspace{-0.4cm}
\plotone{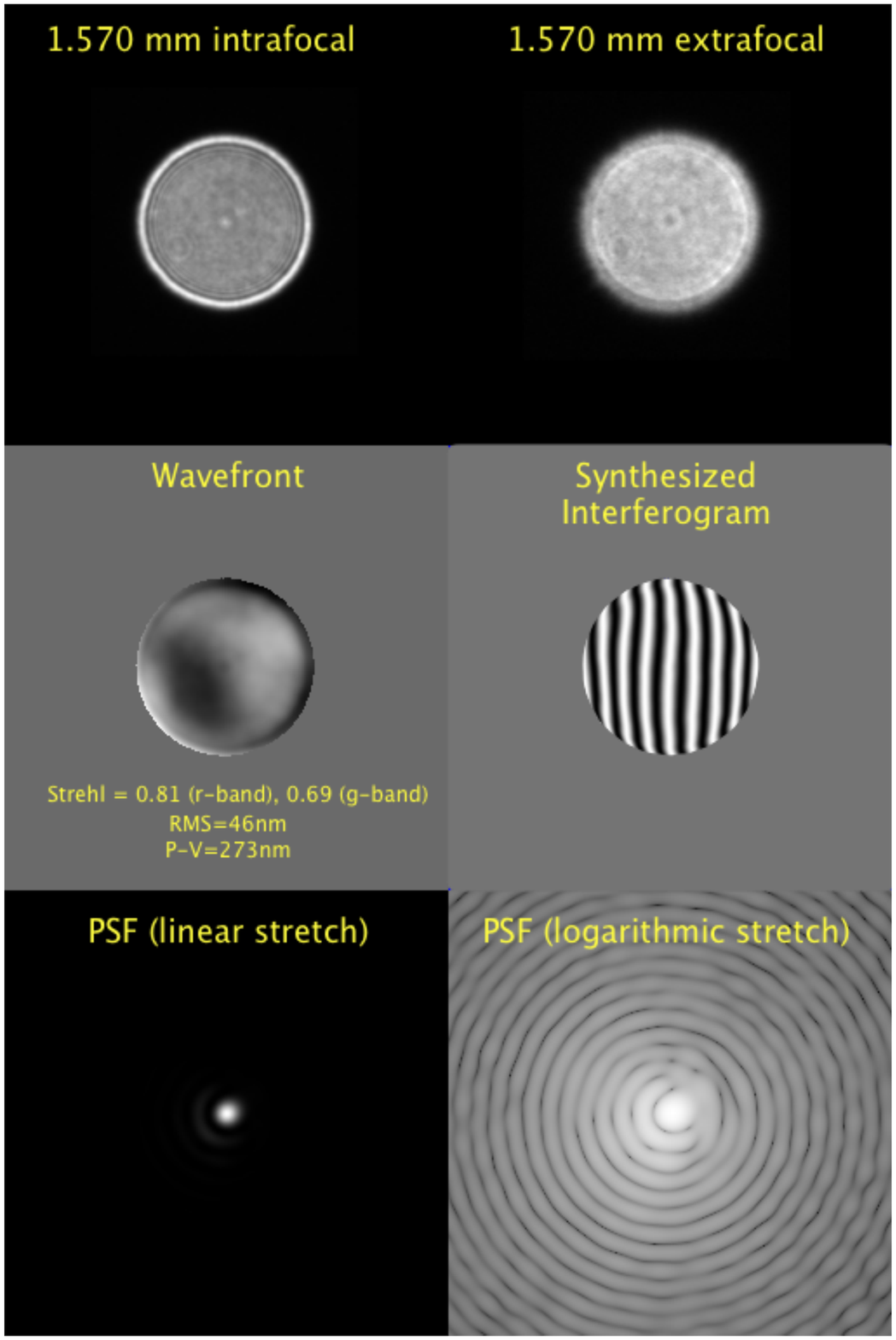}
\vspace{-0.4cm}
\caption{\small
Results from a wavefront curvature analysis of one of the better Canon lenses in the
Dragonfly Telephoto Array. Quantities were obtained using the methodology outlined in
\citet{roddier:93}. All results shown were obtained from two 0.5s integrations of Vega.
[Top row:] intrafocal (left) and extrafocal (right) short-exposure images obtained in
$r$-band. The defocus was 1.574mm (500 digital set points). Note the Fresnel rings are
much softer in the extra-focal image. This is indicative of significant aberrations in the
wavefront. [Middle row:] The computed wavefront (left) and the corresponding synthetic
interferogram (right). The RMS error in the wavefront is 46nm. This gives a Strehl ratio
of 0.81. [Bottom row:] Point-spread function computed from the wavefront in the middle
row. The PSF is shown with linear stretch (left) and logarithmic stretch (right). Note
that the first airy ring is slightly non-uniform but this non-uniformity disappears at
larger angles. See text for details.
\label{lens.plot}}
\end{figure}

Telephoto lenses are generally optimized for minimal chromatic aberration, excellent field
flatness, high illumination over a large area, relatively low distortion, and other
factors. However, they are not usually designed to be diffraction limited at full
aperture. Indeed, as noted earlier, our configuration's 2.85 arcsec pixels greatly
undersample a diffraction limited instrument with the 143mm aperture of a single lens.
Nevertheless, the particular model of Canon lens chosen for use in the Dragonfly Telephoto
Array is reputed to be amongst the sharpest currently available\footnote{Modulation
transfer functions for most Canon lenses are available in Chapter 10 of the {\em Canon EF
Lens Work Book} which can be downloaded from {\tt http://software.canon-europe.com}.}, so
curiosity drove us to explore the ultimate resolution limit of the lenses, even though our
project does not exploit this.  We investigated the on-axis image quality of the lenses
using a curvature wavefront estimator outlined in \citet{roddier:93}. In this procedure,
the wavefront $W$ can be reconstructed from an intra-focal image $I_{\rm in}$ of a bright
star and an extra-focal image $I_{\rm out}$ of the same star by solving a partial
differential equation of the inhomogeneous Poisson form:
\begin{equation}
\nabla^2W \approx k\, {{I_{\rm in} - I_{\rm out}}\over{I_{\rm in} + I_{\rm out}}}.
\end{equation}
\noindent 
subject to the Neumann boundary condition:
\begin{equation}
\partial W/\partial\vec{n} = 0
\end{equation}
\noindent where $k$ is a constant that depends on the degree of defocus and on the focal length of
the lens and $\vec{n}$ is a vector normal to the circular boundary of the projection of
the pupil on the image plane. 

Equation (1) is straightforward to solve using Fourier techniques, and results are
presented for one of the better lenses in the array in Figure~\ref{lens.plot}. The RMS
error in the wavefront is 46nm, corresponding to a Strehl ratio $0.81$ in $r$-band. The
lenses have remarkably good control of chromatic aberration, and the wavefront shape is
similar in $g$-band, corresponding to a formal Strehl ratio of 0.69. Since we estimated
the wavefront from two defocused images (one intrafocal, and one extrafocal), both of
which were affected by seeing, the delivered Strehl from the optics alone is probably
above 0.8 even in $g$-band. Therefore we suspect this lens is diffraction limited at both
wavelengths of interest for our project. However, it is worth noting that the lenses do
show quite significant sample-to-sample variation, with a mean Strehl ratio around 0.4 and
with the poorest lenses having Strehl ratios as low as 0.2. Interestingly, because of our
large pixels, in practice the poorest of our lenses appears to deliver identical
performance to the best lenses when investigated using metrics such as those shown in
Figure~\ref{quality.plot}. The lower Strehl lenses are only identifiable by mapping their
wavefronts, reinforcing our view that the main limiting factor on resolution on our data
is detector sampling and not optical quality (or seeing)\footnote{Note that much higher
Strehl ratios ($>0.95$) are attainable in purpose-built astrographs intended to provide
high-resolution images of astronomical targets. We considered implementing an array of
commercial astrographs, but after looking carefully at the trade-offs involved it was
concluded that an array of telephoto lenses was the better choice in our case.}. For our
purposes the most important characteristics of the lenses are defined by their scattering
and ghosting properties, the investigation of which we turn to next.

\subsection{Scattering and ghosting}

\begin{figure*}[htbp]
\epsscale{2.0}
\plotone{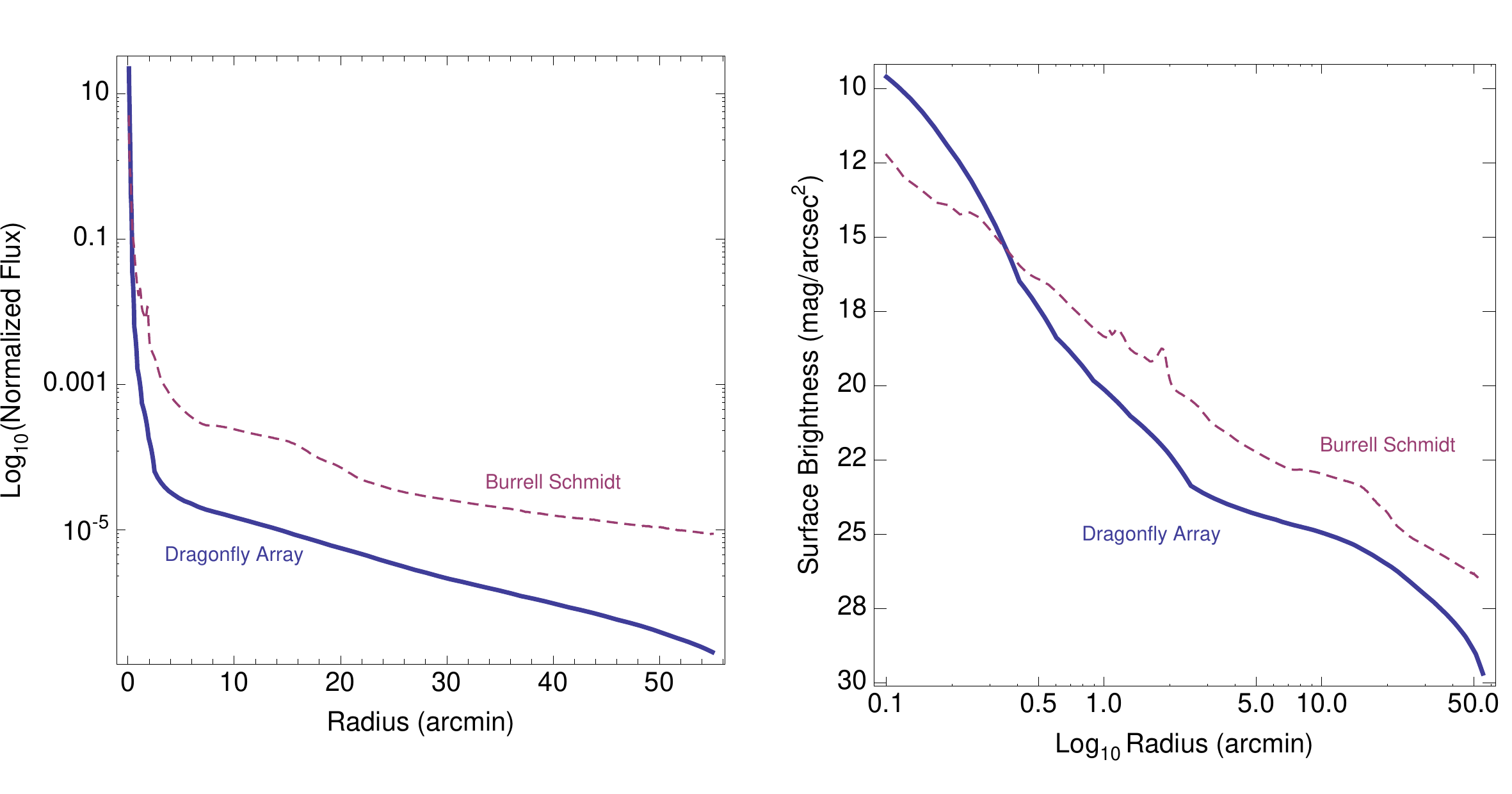}
\caption{\small
Comparison of the wide-angle halo point-spread function of the Canon telephoto lenses used
in the Dragonfly Telephoto Array with the corresponding point-spread function of the
0.9\,m Burrell Schmidt telescope on Kitt Peak (Slater et al.\ 2009). The Burrell Schmidt
is optimized for low surface brightness imaging and has produced a well-known ultra-deep
image of the Virgo cluster (Mihos et al.\ 2005). The minimum radius plotted is 0.1 arcmin,
and in both cases $\gtrsim90\%$ of the total stellar flux is interior to this, so only a
small fraction of a star's total light is contained within the wide-angle halo (mainly due
to scattering and internal reflections) shown in these plots. [Left Panel:] Flux as a
function of linear radius, after normalizing both profiles to contain identical total
fluxes. Note that in large part thanks to nano-fabricated anti-reflection coatings on some
of its elements, the Canon lenses have a factor of $5-10$ less halo light at radii $>5$
arcmin. [Right Panel:] The $r$-band AB mag/arcsec$^2$ surface brightness profile of Vega
as a function of logarithmic radius for the Dragonfly Array and for the Burrell Schmidt
telescope. See text for details.
\label{profile.plot}}
\end{figure*}

To examine the scattering and ghosting properties of the Dragonfly Telephoto Array's
lenses we imaged Vega and used this to determine the radially-averaged point spread
function of one of our lenses out to angular radii around one degree. We wanted to explore
contrast ratios of around $10^8$:1, which is far more than can be accommodated with a
single integration. Therefore the star was imaged at a range of integration times and a
series of sub-profiles was constructed. These sub-profiles were stitched together to
create the complete profile shown in Figure~\ref{profile.plot}. This figure also shows the
published point-spread function from the 0.9\,m Burrell Schmidt telescope on Kitt Peak
(Slater et al.\ 2009). The minimum radius plotted is 0.1 arcmin, and $\gtrsim90\%$ of the
total stellar flux is interior to this, so only a small fraction of a star's total light
is contained within the wide-angle halo (mainly due to scattering and internal
reflections) shown in the figure. The left hand panel of Figure~\ref{profile.plot} shows
flux as a function of linear radius, after normalizing both the Dragonfly and Burrell
profiles so they contain identical total fluxes. The right hand panel shows the $r$-band
AB mag/arcsec$^2$ surface brightness profile of Vega as a function of logarithmic radius.
The former is based on direct measurement, but the latter is an estimate based on the
profile of Arcturus presented in Slater et al. (2009), who imaged this star in the
Washington $M$-band. In transforming to the Vega profile we accounted for the relative
brightnesses of the two stars and the $M-r$ colour of Arcturus.

The Canon lenses in the Dragonfly Telephoto Array show a factor of $5-10$ less scattered
light at radii $>5$ arcmin than the Burrell Schmidt. The point spread function from the
lenses declines more steeply than that of the Burrell Schmidt, and shows fewer bumps from
diffraction effects and ghosts. The Burrell Schmidt is a superb instrument that is highly
optimized for low surface brightness imaging and it has produced a beautiful and
well-known ultra-deep image of the Virgo cluster (Mihos et al.\ 2005). Therefore we
consider the performance of the commercial lens exhibited in Figure~\ref{profile.plot} to
be quite impressive.

\begin{figure}[htbp]
\epsscale{1.0}
\plotone{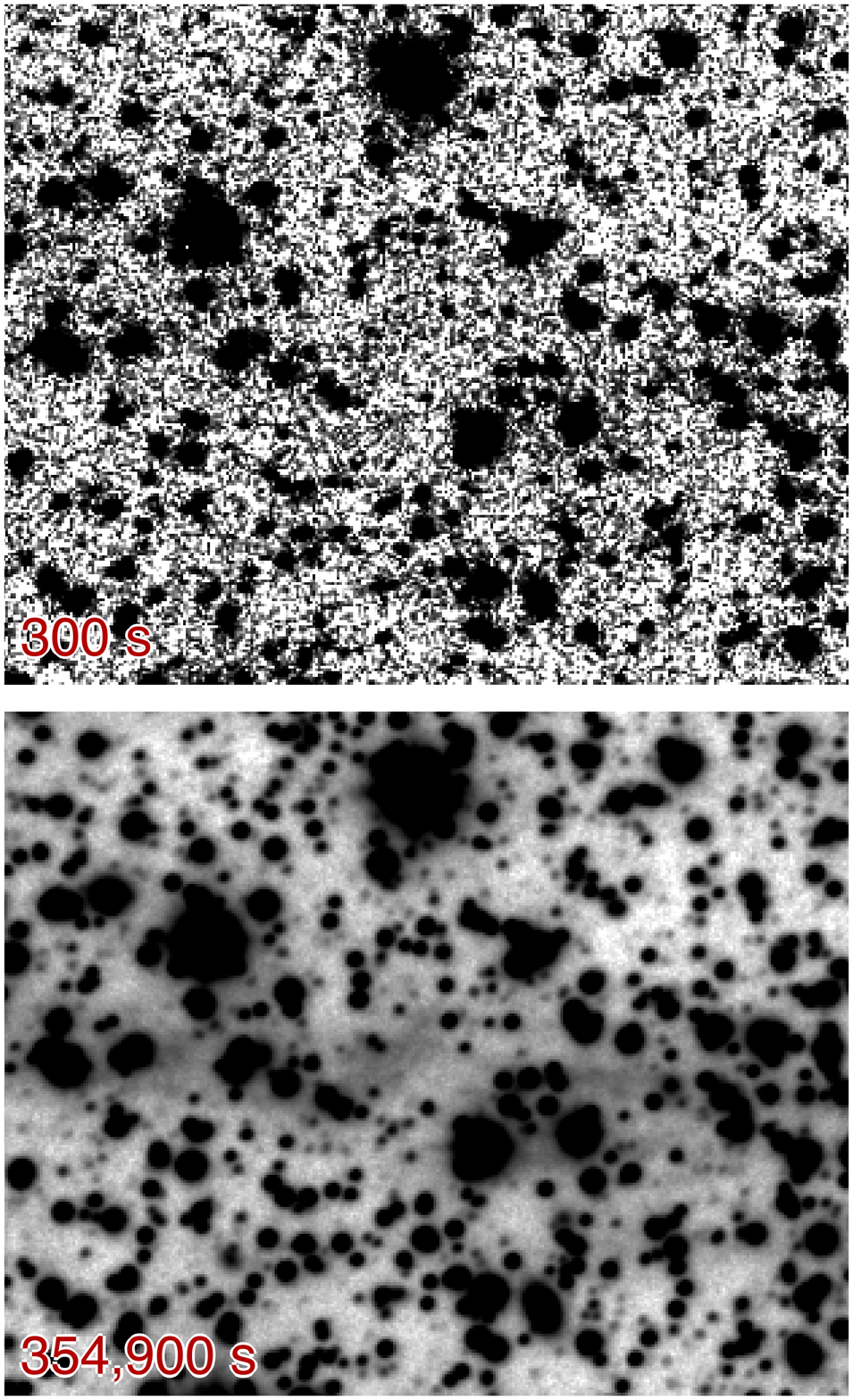}
\caption{\small
Illustration of the excellent control of scattered light in the Dragonfly Telephoto Array
PSF. The image is a $13.1' \times 10.5'$ area of crowded sky near the Galactic plane. The
top panel shows an individual 300\,s exposure obtained with a single telephoto lens. The
bottom panel shows a combination of many exposures in order to provide an equivalent of a
$1183 \times$ longer exposure time. Nevertheless, the stellar halos do not appear much
bigger in the bottom image. Note the faint low surface brightness cirrus that is apparent
in the combined image. The data in this figure was obtained in a filterless `white light'
configuration and we cannot calibrate the surface brightness of the structure shown, but
based upon carefully calibrated data through $g$-band and $r$-band observations of M101
(van Dokkum et al. 2013, submitted) it is clear that some of the structures shown are at
surface brightnesses below 30~\sb.
\label{depth_compare.plot}}
\end{figure}

What new capabilities for low-surface brightness imaging are opened up by optics that
deliver the performance shown in Figure~\ref{profile.plot}? \citet{slater:09} showed that
even at the high galactic latitude of the Virgo cluster ($b\sim75^\circ$), using the
optimized Burrell Schmidt, over 50\% of pixels are contaminated by scattered light from
faint stars at $\mu=29$~\sb. Figure~\ref{profile.plot} suggests that equivalent
performance can be obtained around two magnitudes fainter using the Dragonfly Telephoto
Array. The ramifications of this are displayed in Figure~\ref{depth_compare.plot}. The top
panel of this figure shows a small area of an individual 300\,s exposure of a crowded
galactic field obtained using a single telephoto lens. The bottom panel shows the
corresponding area imaged using a combination of many exposures from separate lenses to
provide the equivalent of $1183 \times$ longer exposure time. The images are stretched to
a similar dynamic range. This figure makes two important points. Firstly, in spite of the
integration time being over 1000x longer in the bottom panel, the stellar halos are not
much bigger. We have not yet encountered systematic errors limiting the usefulness of very
long integration times (up to $\sim 100$ hours) with the Dragonfly Telephoto Array. The
second important point is that at very low surface brightness levels even apparently empty
patches of sky can be riddled with cirrus. The need for foreground structure avoidance (or
removal) will play an important role in defining the strategy for surveys intended to
explore faint galactic structures. We will have much more to say about this topic in
future papers.

\begin{figure}[htbp]
\epsscale{0.9}
\plotone{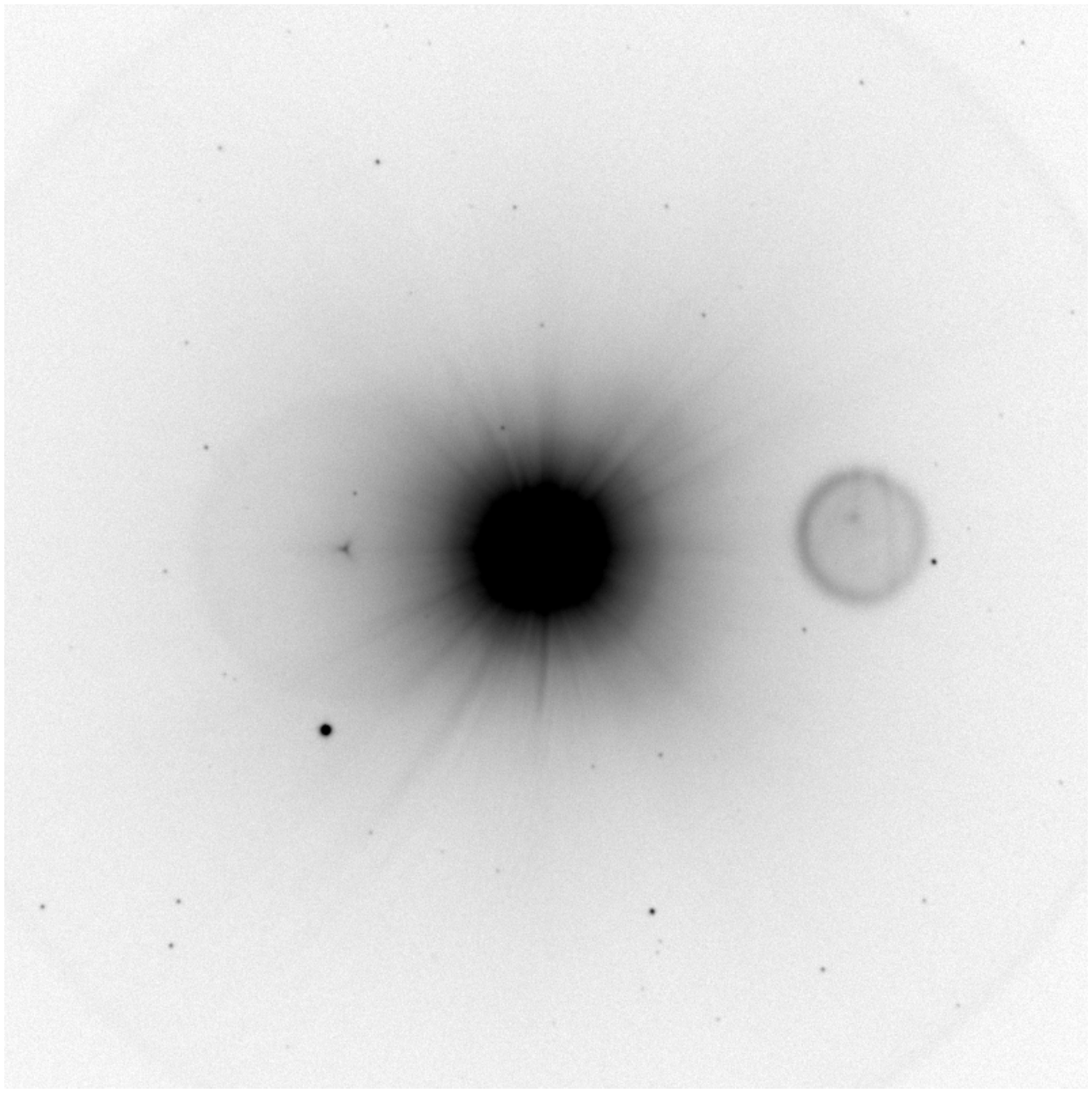}
\caption{\small
A white-light image of the planet Venus, obtained with the one of the Canon 400/2.8 EF
IS~II telephoto lenses of the Dragonfly Telephoto Array on Nov. 4, 2012. The image spans
$50' \times 50'$ is displayed with a logarithmic stretch. The PSF is well behaved out to
large radius and shows no strong diffraction spikes and relatively few large ghosts or
other asymmetric features. On the date the image was taken Venus was at magnitude $-4$,
and is highly saturated on the frame. However the bright star to the bottom-left of Venus
is unsaturated. This is the $\sim7$th magnitude F5 star SAO 138876 ($m_B = 7.42~{\rm mag},
m_V =6.95~{\rm mag}$), which can be used as a calibrator. Using this star as a reference,
the highest surface brightness ghost has an integrated flux $\sim 9$ mag fainter than that
of Venus, and contains only $\sim 0.025\%$ of its light.
\label{venus.plot}}
\end{figure}

We conclude this section by noting that, in spite of their excellent performance in terms
of ghosting and scattered light control, the lenses in our array still have room for some
improvement. Figure~\ref{venus.plot} shows a $1^\circ~{\rm x}~1^\circ$ subset of an image
of Venus imaged near maximum brightness (apparent mag $\sim -4$) obtained with one of the
telephoto lenses. Ghosts and wide-angle diffraction effects in this image are seen at a
much lower level than in images obtained with reflecting telescopes (for example, compare
this figure with the images of Arcturus shown in Slater et al. 2009), but they are not
completely absent. We are uncertain of the origin of the low-level spikes seen in
Figure~\ref{venus.plot}. We tentatively attribute them to diffraction from the blades of
the internal iris in the lens which enters into the edge of the pupil when a target is not
perfectly on axis, but it is also possible that at this level there are some contributions
to diffraction from striae in the glasses or lattice imperfections in the crystalline
optical elements used by the lenses. We attribute the circular ghosts to the fact that not
all optical elements in the lens use the nanostructure coatings and also the coatings
themselves are not perfect. In spite of this, overall performance is quite impressive: the
highest surface brightness ghost in the image (to the immediate right of Venus) contains
only $\sim0.025\%$ of the light from the planet. Our strategy for dealing with these
features is to take advantage of the fact that their position depends on the optical path
through the lenses, so we can offset individual cameras slightly and average the features
out when images are combined. This takes advantage of the redundancy at the heart of the
array design, and this aspect of the instrument concept also plays a key role in
mitigating the effects of imperfect flat fielding and time-dependent additive atmospheric
contributions to the images. In practice, summed images are flat to better than 0.1\% of
the sky level. Readers interested in the methodology for the data reduction procedure
currently used with the Dragonfly Telephoto Array data are referred to van Dokkum, Abraham
\& Merritt (2013).

\section{Sensitivity}

\begin{figure*}[htbp]
\epsscale{2.2}
\plotone{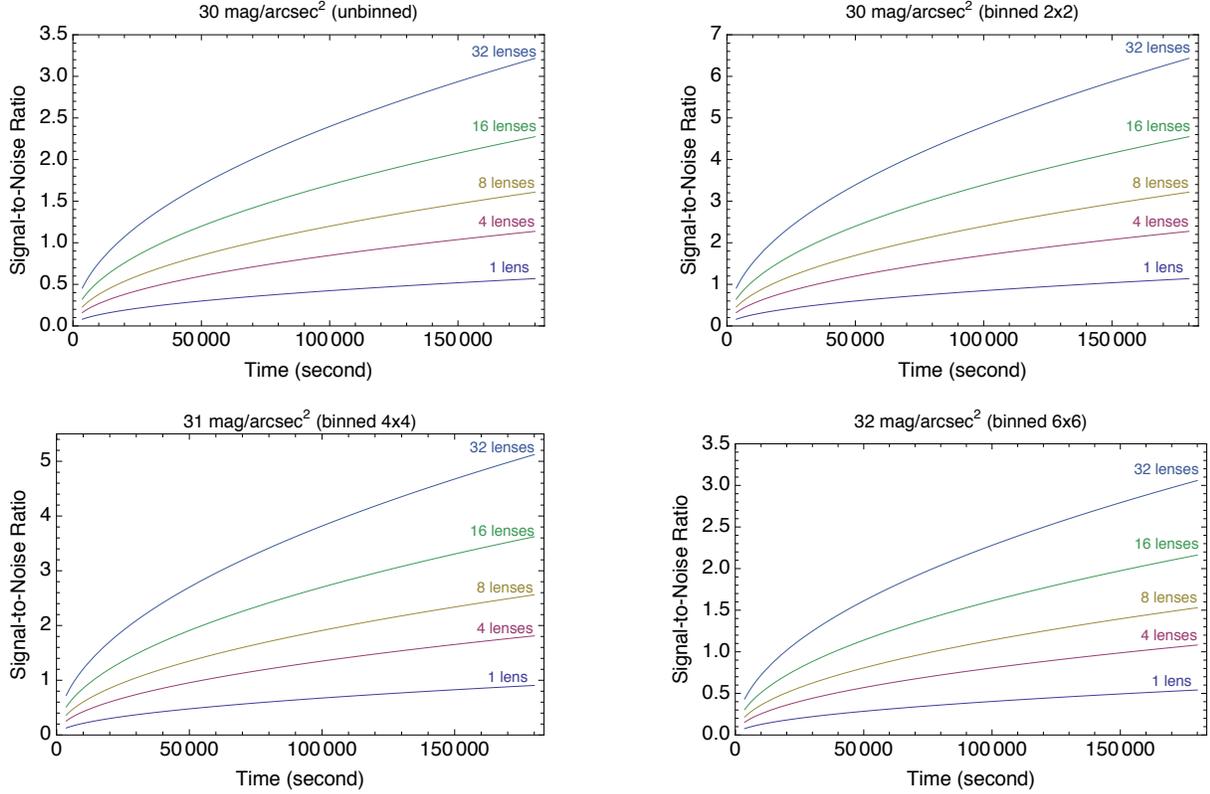}
\caption{
SDSS $g$-band signal-to-noise ratio as a function of integration time  for various
configurations of the Dragonfly Telephoto Array. Individual panels correspond to targets
with surface brightness and binning factors of (top-left:) 30 mag/arcsec$^2$ with no
binning; (top-right:) 30 mag/arcsec$^2$ binned 2x2; (bottom-left:) 31 mag/arcsec$^2$
binned 4x4; (bottom-right:) 32 mag/arcsec$^2$ binned 6x6. different surface brightness and
binning factor. The lines on individual panels correspond to the performance of 32-lens,
16-lens, 8-lens (current configuration), 4-lens, and 1-lens systems. All panels assume a
$g$-band sky surface brightness of 22.5 mag/arcsec$^2$, which is the darkest sky expected
at the New Mexico Skies telescope hosting facility at solar minimum. 
}
\label{snr.plot}
\end{figure*}

The rate of detected photons per pixel, $\Phi$,  from an object with surface brightness
$S_\lambda(\lambda)$ is given by:
\begin{equation}
\Phi = a\ \left({D \over L}\right)^2  \int  {E(\lambda) S_\lambda(\lambda) \over h \nu}  e^{-k(\lambda)\sec Z} d\lambda
\label{rate.equation}
\end{equation}

\noindent where $a$ is the area of a pixel, $D$ is the effective aperture, $L$ is the
system focal length, $E(\lambda)$ is the total system efficiency, $\sec Z$ is the air mass
and $k(\lambda)$ is the extinction coefficient of the atmosphere. We can convert this
equation into a more immediately useful form by: (i) ignoring extinction; (ii) noting that
(D/L) is the inverse of the focal ratio $f$; (iii)  replacing the functions inside the
integral by their mean values within the filter bandpass $\Delta\lambda$; and (iv)
converting the frequency in the denominator to an equivalent wavelength. This gives:
\begin{equation}
\Phi \approx a\ f^{-2}\  {\bar{E} \bar{S}_\lambda \over h c} \bar{\lambda}\ \Delta\lambda
\label{simplified.rate.equation}
\end{equation}

\noindent which reinforces the point noted earlier that the detected photon rate per pixel
varies inversely with the square of the focal ratio. This is of course the central reason
why the Dragonfly Telephoto Array (effectively an $f/1$ system in its current
configuration) is quite efficient\footnote{Of course
equation~\ref{simplified.rate.equation} also shows that a slow optical system can increase
its detected photon rate by detector binning. For survey instruments there is generally an
\'etendue conservation trade to be made between aperture, focal ratio and desired spatial
resolution.}.

The surface brightness in equation~\ref{simplified.rate.equation} can be cast into units
of mag/arcsec$^2$ by noting that $\bar{S}_\lambda = \bar{F}_0\times 10^{-0.4 \mu}$, where
$\bar{F}_0$ is the flux zero point for the filter and $\mu$ is the mean surface brightness
in magnitude units in the bandpass. In the AB system the zeroth magnitude calibrator is
defined in frequency space as 3631 Jy in all bands, so after conversion back to wavelength
space $\bar{F}_0 = (3631\ {\rm Jy/arcsec^2})\times c/\bar{\lambda}^2$ where
$\bar{\lambda}$ is the effective wavelength of the filter. Substituting this into
equation~\ref{simplified.rate.equation} and rearranging, we obtain:
\begin{equation}
\Phi \approx {a\ f^{-2}\ \bar{E}\ (3631\ {\rm Jy/arcsec^2})\ 10^{-0.4 \mu} \over h}   \ \left(\Delta\lambda \over \bar{\lambda}\right)
\label{mag.rate.equation}
\end{equation}

Before inserting numerical values into this equation it is important to
note two things:

\begin{enumerate}
\item  The transmission curve of an interference filter is a function of the beam speed.
Because the beam of the telephoto lenses used in the Dragonfly Array is quite fast, some
care needs to be taken before using manufacturer's scans obtained using a collimated beam
to define $\bar{\lambda}$ and $\Delta\lambda$. The change in the transmission curve as a
function of beam speed is a function of the specific recipe used to construct the
dielectric layer stack in the filter. However, in general, as the beam speed increases the
transmission curve shifts very slightly blueward and the peak throughput drops.\footnote{An
approximate expression for the shift in central wavelength is $\bar{\lambda}_{\rm eff} =
\bar{\lambda}(1-{\beta^2\over 4 n^2})$ where $\bar{\lambda}_{\rm eff}$ is the shifted
central wavelength, $n$ is the average index of refraction for the multiple layers of
dielectric in the stack, and $\beta$ is the the cone angle (the maximum incidence angle of
light in the filter). The cone angle is related to focal ratio by $\sin{\beta} \sim
1/(2f)$. The loss in peak transmission is more difficult to model and we are unaware of
any simple `rule of thumb' approximation for this.} The latter effect is seen fairly clearly in
Figure~2 when comparing the Astrodon filter measurements made with a collimated beam to
CFHT MegaCam measurements made with an $f/8$ beam. We do not (yet) have transmission
curves measured at $f/2.8$, but the CFHT MegaCam filter transmission curves are likely to
be a closer approximation to the actual transmission of the filters in our configuration
than are the Astrodon curves.
We therefore use the CFHT MegaCam throughput curves in our calculations of the
throughput of the Canon lenses given below.
 
\item There is some ambiguity in the literature regarding the definitions of filter
effective wavelengths and bandpasses. In the present context, we define the effective
wavelength to be the mean wavelength: \begin{equation} \bar{\lambda} = \int \lambda\ 
\phi(\lambda)\  d\lambda \end{equation} where $\phi$ is the transmission curve normalized
to unit area. We define the bandpass $\Delta\lambda$ to mean the full-width at half
maximum obtained by approximating the normalized transmission curve by a gaussian, which
gives: \begin{equation} (\Delta\lambda)^2 = 8 \ln2 \int (\lambda - \bar{\lambda})^2\ 
\phi(\lambda)\  d\lambda. \end{equation} \noindent With these definitions,  the SDSS $g$
and $r$-band transmission curves based on CFHT MegaCam measurements shown in
Figure~\ref{sens.plot} give $\bar{\lambda} = 485.7$nm and $\Delta\lambda = 100.2$nm
($g$-band) and $\bar{\lambda} = 627.3$nm and $\Delta\lambda = 84.2$nm ($r$-band).
\end{enumerate}

Inserting numerical values into equation (5) we obtain the following count rates, 
corresponding to $\mu=0\ {\rm mag}/{\rm arcsec}^2$ at the top of the
atmosphere in $g$ and $r$-band,
respectively:
\begin{eqnarray}
\Phi_g & = & 1.79\times 10^9\ N\ \bar{E}\  {\rm photon/pix/s} \\
\Phi_r & = & 1.16\times 10^9\ N\ \bar{E}\  {\rm photon/pix/s}.
\end{eqnarray}
\noindent In these expressions $N$ is the number of lenses in the array, and $\bar{E}$ is
the mean efficiency. These equations can be used to compute the efficiency of the lenses
from measured instrumental zero points and appropriate corrections for extinction.

Comparing the count rates in equations (8) and (9) with observations made on August 1,
2013, we find that the overall efficiency of the Dragonfly Array was 34\% in $g$-band and
35\% in $r$-band, with individual lenses varying by 1--2 percent around these
values\footnote{We incorporated 0.39~mag of $g$-band extinction and 0.21~mag of $r$-band
extinction into the estimates. These values were calculated by summing Rayleigh
scattering, ozone absorption and aerosol pollutant terms. We note that the aerosol
(microscopic particulate) term is often neglected, but it is often important; for example,
in summer it is often the dominant term for observatories located in the Southwestern USA
because of forest fires. It is straightforward to determine the needed correction for
particulates from aerosol optical depth maps produced by the Moderate Resolution Imaging
Spectroradiometer (MODIS) on the NASA Terra/EOS-AM satellite. At the present time, nightly
MODIS maps are available here: http://www.star.nesdis.noaa.gov/smcd/spb/aq/index.php.}.
These observations were made several months after cleaning, and in our experience freshly
cleaned lenses improve upon these numbers by several percent. These observed efficiencies
should be compared with the maximum efficiency achievable (43\% in both $g$ and $r$-bands)
based on the throughput curves for the filters and the quantum efficiency of the detector
shown in Figure~\ref{sens.plot}. We conclude that a lower limit for the overall
transmittance of freshly cleaned\footnote{An interesting advantage of an all-refractive
design is that it is less sensitive to dirt. A dirty objective decreases 
throughput without much of an an impact on contrast because the dominant
scattering mode is backwards and out of the optical path. On the other hand,
scattering from a reflective
objective is more pernicious, because it scatters light back into the beam. In other words, dust on
reflective surfaces not only dims the view, it also reduces contrast by increasing the sky
background.} lenses is 85\%, and the weak link in the overall efficiency of the system at
present is the relatively low quantum efficiency of the commercial CCD cameras used.
Improving these is an obvious upgrade for the array.

Unlike most imaging systems, the Dragonfly Telephoto Array is designed to grow in
effective aperture with time as additional lenses are added to the system. An attempt to
translate the performance of the Dragonfly Telephoto Array into signal-to-noise ratio as a
function of source surface brightness, integration time, array size, and binning factor is
shown in Figure~\ref{snr.plot}. This figure makes two important points. The first of these
is that, because the system is not limited by systematic errors, integration times in
excess of 100ks are entirely realistic. The second is that even with the array in its
present configuration, surface brightness levels well below 30 mag/arcsec$^2$ are easily
achievable. By trading off resolution, surface brightnesses below 32 mag/arcsec$^2$ are
realistically achievable by binning images with the current array. Alternatively, one
could simply upgrade the array to incorporate more lenses. In any case, undertaking a
survey of around 100 galaxies to the surface brightness limits needed to test the
predictions of galaxy formation models is feasible, but will require the present array to
be operated in a survey mode for several years. We have embarked on such a survey, and the
first results from this campaign are presented in van Dokkum, Abraham \& Merritt 2013
(submitted).

\section{Summary}

The Dragonfly Telephoto Array is an innovative instrument concept whose goal is to open up
a new observational regime: ultra low surface brightness imaging at visible wavelengths.
The array is comprised of multiple commercial 400mm $f/2.8$ telephoto lenses which have
high performance sub-wavelength nano-fabricated optical coatings designed to minimize
scattered light and ghosting. The array is presently comprised of eight lenses, though we
anticipate adding additional lenses in the future. In this paper we have outlined the
basic concept and reported on the performance of the 8-lens array, which is similar to
that of a 0.4\,m aperture $f/1.0$ refractor with a $2.6^{\circ}\times 1.9^{\circ}$ field
of view. The imager has scattering and ghosting properties an order of magnitude better
than those of the best reflectors optimized for low surface brightness imaging. Harnessing
this new capability, the Dragonfly Telephoto Array is now executing a fully-automated
multi-year imaging survey of a  sample of nearby galaxies in order to undertake the first
census of ultra-faint substructures in galaxies in the nearby Universe. The detection of
structures at these surface brightness levels may hold the key to solving the `missing
substructure' and `missing satellite' problems of conventional hierarchical galaxy
formation models. 

\begin{acknowledgements}
\noindent{\em Acknowledgments}
\\
We are indebted to the Dunlap Institute of the University of Toronto, Yale University, and
NSERC for financial support of this project. We particularly thank Peter Martin and James
Graham, the acting and former directors of the Dunlap Institute at the University of
Toronto, for their moral and financial support and excellent technical advice. PhD
students Jielai Zhang (Toronto) and Allison Merritt (Yale) deserve many thanks for
engineering work, for debugging, and for late nights observing. Dana Simard (Queens
University) is thanked for her work in developing the Dragonfly Web App.

Ren\'e Doyon and Robert Lamontagne of the Universit\'e de Montr\'eal
deserve our thanks for allowing us to undertake the initial testing of the
lenses used in this project at the Observatoire du Mont M\'egantic. 

We are grateful for all the help provided by the staff at New Mexico Skies, and for their
outstanding professionalism in support of this project. We particularly thank Mike Rice
and Lynn Rice, owners of facility, and site engineers Randy Reimers and Grady Owens, for
superb support and for innumerable late night fixes to the Dragonfly Telephoto Array.
Thanks are also due to Erik Widding of Birger Engineering and Don Goldman of Astrodon
Imaging for supplying us with customized focus controllers and filters for this project.
We are also grateful to Software Bisque for supplying us with an excellent Paramount ME II
mount on short order, and to Matthew Bisque for technical assistance.

This project emerged as a result of a bet\footnote{RGA lost.} made by RGA and PGvD at the
Mount Everest Nepalese Restaurant on Bloor Street in Toronto. We thank the United
Breweries Group of Bangalore for providing us with inspiration on the night in question.
\end{acknowledgements}

\bigskip

\appendix
\noindent{\bf Appendix 1 -- Example of a nightly control script}

\begin{verbatim}
#!/bin/sh

# Edit this script in the afternoon, set it running, then get your data tomorrow.
#
# To observe a target specified by name, use the 'auto_observe' command below. This pretty
# much does everything for you. To be more specific: 'auto_observe' monitors the state of the 
# observatory and only integrates if the weather is good enough for the
# roof to be open. If the weather is not good enough it simply sleeps for for the equivalent 
# exposure time and then continues at the next step in the sequence, checking the weather
# as it goes along. Once observing commences, the script focuses all of the
# cameras then executes a 9-point dither pattern with 600s integrations. Guide stars
# are acquired as needed. A dark frame is obtained at the end of the sequence.
# Focus is adjusted as needed. At the end of the sequence all-sky camera
# images, image quality estimates, and other log information is emailed to the 
# Dragonfly account.

######### Startup settings - do not edit the next set of lines #########

source "/usr/local/bin/df_library.sh"   # Load library
all_set_path                            # Automatically set data location
df_sleep_until_twilight                 # Pause until twilight
df_startup                              # Activate system with best-guess focus positions
all_regulate -10                        # Set CCD temperature
auto_flats                              # Automatic flats with 5000-15000 ADU in r'-band

######### EDIT BELOW HERE #########

sleep_until 0256                        # Set to end of nautical twilight in UTC

# Observe M101 tonight. The moon rises in about 6h so stop then.

auto_observe "M 101"                    # This line takes about 2h to execute
auto_observe "M 101"
auto_observe "M 101"

######### DO NOT EDIT BELOW HERE #########

# Done observing. Shut down gracefully.
if [ -e "INTERRUPT.txt" ]
then
  exit
else
  mount park
  power 12V off
fi
\end{verbatim}

\vfill\eject

\bibliographystyle{apj}
\bibliography{ms}

\begin{thebibliography}{25}
\expandafter\ifx\csname natexlab\endcsname\relax\def\natexlab#1{#1}\fi

\bibitem[{{Atkinson} {et~al.}(2013){Atkinson}, {Abraham}, \&
  {Ferguson}}]{atkinson:13}
{Atkinson}, A.~M., {Abraham}, R.~G., \& {Ferguson}, A.~M.~N. 2013, \apj, 765,
  28

\bibitem[{{Barton} \& {Thompson}(1997)}]{barton:97}
{Barton}, I.~J. \& {Thompson}, L.~A. 1997, \aj, 114, 655

\bibitem[{{Belokurov} {et~al.}(2007){Belokurov}, {Evans}, {Irwin},
  {Lynden-Bell}, {Yanny}, {Vidrih}, {Gilmore}, {Seabroke}, {Zucker},
  {Wilkinson}, {Hewett}, {Bramich}, {Fellhauer}, {Newberg}, {Wyse}, {Beers},
  {Bell}, {Barentine}, {Brinkmann}, {Cole}, {Pan}, \& {York}}]{belokurov:07}
{Belokurov}, V., {Evans}, N.~W., {Irwin}, M.~J., {Lynden-Bell}, D., {Yanny},
  B., {Vidrih}, S., {Gilmore}, G., {Seabroke}, G., {Zucker}, D.~B.,
  {Wilkinson}, M.~I., {Hewett}, P.~C., {Bramich}, D.~M., {Fellhauer}, M.,
  {Newberg}, H.~J., {Wyse}, R.~F.~G., {Beers}, T.~C., {Bell}, E.~F.,
  {Barentine}, J.~C., {Brinkmann}, J., {Cole}, N., {Pan}, K., \& {York}, D.~G.
  2007, \apj, 658, 337

\bibitem[{{Bertin} \& {Arnouts}(1996)}]{bertin:96}
{Bertin}, E. \& {Arnouts}, S. 1996, \aaps, 117, 393

\bibitem[{{Cooper} {et~al.}(2010){Cooper}, {Cole}, {Frenk}, {White}, {Helly},
  {Benson}, {De Lucia}, {Helmi}, {Jenkins}, {Navarro}, {Springel}, \&
  {Wang}}]{cooper:10}
{Cooper}, A.~P., {Cole}, S., {Frenk}, C.~S., {White}, S.~D.~M., {Helly}, J.,
  {Benson}, A.~J., {De Lucia}, G., {Helmi}, A., {Jenkins}, A., {Navarro},
  J.~F., {Springel}, V., \& {Wang}, J. 2010, \mnras, 406, 744

\bibitem[{{Cooper} {et~al.}(2013){Cooper}, {D'Souza}, {Kauffmann}, {Wang},
  {Boylan-Kolchin}, {Guo}, {Frenk}, \& {White}}]{cooper:13}
{Cooper}, A.~P., {D'Souza}, R., {Kauffmann}, G., {Wang}, J., {Boylan-Kolchin},
  M., {Guo}, Q., {Frenk}, C.~S., \& {White}, S.~D.~M. 2013, ArXiv e-prints

\bibitem[{{Dubinski} {et~al.}(1996){Dubinski}, {Mihos}, \&
  {Hernquist}}]{dubinski:96}
{Dubinski}, J., {Mihos}, J.~C., \& {Hernquist}, L. 1996, \apj, 462, 576

\bibitem[{{Fry} {et~al.}(1999){Fry}, {Morrison}, {Harding}, \&
  {Boroson}}]{fry:99}
{Fry}, A.~M., {Morrison}, H.~L., {Harding}, P., \& {Boroson}, T.~A. 1999, \aj,
  118, 1209

\bibitem[{{Ibata} {et~al.}(2001){Ibata}, {Irwin}, {Lewis}, {Ferguson}, \&
  {Tanvir}}]{ibatam31:01}
{Ibata}, R., {Irwin}, M., {Lewis}, G., {Ferguson}, A.~M.~N., \& {Tanvir}, N.
  2001, \nat, 412, 49

\bibitem[{{Johnston} {et~al.}(2008){Johnston}, {Bullock}, {Sharma}, {Font},
  {Robertson}, \& {Leitner}}]{johnston:08}
{Johnston}, K.~V., {Bullock}, J.~S., {Sharma}, S., {Font}, A., {Robertson},
  B.~E., \& {Leitner}, S.~N. 2008, \apj, 689, 936

\bibitem[{{Kazantzidis} {et~al.}(2008){Kazantzidis}, {Bullock}, {Zentner},
  {Kravtsov}, \& {Moustakas}}]{kazantzidis:08}
{Kazantzidis}, S., {Bullock}, J.~S., {Zentner}, A.~R., {Kravtsov}, A.~V., \&
  {Moustakas}, L.~A. 2008, \apj, 688, 254

\bibitem[{{Kormendy} \& {Bahcall}(1974)}]{kormendy:74}
{Kormendy}, J. \& {Bahcall}, J.~N. 1974, \aj, 79, 671

\bibitem[{{Mart{\'{\i}}nez-Delgado} {et~al.}(2010){Mart{\'{\i}}nez-Delgado},
  {Gabany}, {Crawford}, {Zibetti}, {Majewski}, {Rix}, {Fliri},
  {Carballo-Bello}, {Bardalez-Gagliuffi}, {Pe{\~n}arrubia}, {Chonis}, {Madore},
  {Trujillo}, {Schirmer}, \& {McDavid}}]{martinez:10}
{Mart{\'{\i}}nez-Delgado}, D., {Gabany}, R.~J., {Crawford}, K., {Zibetti}, S.,
  {Majewski}, S.~R., {Rix}, H.-W., {Fliri}, J., {Carballo-Bello}, J.~A.,
  {Bardalez-Gagliuffi}, D.~C., {Pe{\~n}arrubia}, J., {Chonis}, T.~S., {Madore},
  B., {Trujillo}, I., {Schirmer}, M., \& {McDavid}, D.~A. 2010, \aj, 140, 962

\bibitem[{{McConnachie} {et~al.}(2009){McConnachie}, {Irwin}, {Ibata},
  {Dubinski}, {Widrow}, {Martin}, {C{\^o}t{\'e}}, {Dotter}, {Navarro},
  {Ferguson}, {Puzia}, {Lewis}, {Babul}, {Barmby}, {Bienaym{\'e}}, {Chapman},
  {Cockcroft}, {Collins}, {Fardal}, {Harris}, {Huxor}, {Mackey},
  {Pe{\~n}arrubia}, {Rich}, {Richer}, {Siebert}, {Tanvir}, {Valls-Gabaud}, \&
  {Venn}}]{mcconnachie:09}
{McConnachie}, A.~W., {Irwin}, M.~J., {Ibata}, R.~A., {Dubinski}, J., {Widrow},
  L.~M., {Martin}, N.~F., {C{\^o}t{\'e}}, P., {Dotter}, A.~L., {Navarro},
  J.~F., {Ferguson}, A.~M.~N., {Puzia}, T.~H., {Lewis}, G.~F., {Babul}, A.,
  {Barmby}, P., {Bienaym{\'e}}, O., {Chapman}, S.~C., {Cockcroft}, R.,
  {Collins}, M.~L.~M., {Fardal}, M.~A., {Harris}, W.~E., {Huxor}, A., {Mackey},
  A.~D., {Pe{\~n}arrubia}, J., {Rich}, R.~M., {Richer}, H.~B., {Siebert}, A.,
  {Tanvir}, N., {Valls-Gabaud}, D., \& {Venn}, K.~A. 2009, \nat, 461, 66

\bibitem[{{Mihos} {et~al.}(2005){Mihos}, {Harding}, {Feldmeier}, \&
  {Morrison}}]{mihos:05}
{Mihos}, J.~C., {Harding}, P., {Feldmeier}, J., \& {Morrison}, H. 2005, \apjl,
  631, L41

\bibitem[{{Mihos} {et~al.}(2013){Mihos}, {Harding}, {Spengler}, {Rudick}, \&
  {Feldmeier}}]{mihos:13}
{Mihos}, J.~C., {Harding}, P., {Spengler}, C.~E., {Rudick}, C.~S., \&
  {Feldmeier}, J.~J. 2013, \apj, 762, 82

\bibitem[{{Moore} {et~al.}(1999){Moore}, {Lake}, {Quinn}, \&
  {Stadel}}]{moore:99}
{Moore}, B., {Lake}, G., {Quinn}, T., \& {Stadel}, J. 1999, \mnras, 304, 465

\bibitem[{{Naab} {et~al.}(2007){Naab}, {Johansson}, {Ostriker}, \&
  {Efstathiou}}]{naab:07}
{Naab}, T., {Johansson}, P.~H., {Ostriker}, J.~P., \& {Efstathiou}, G. 2007,
  \apj, 658, 710

\bibitem[{{Nelson} {et~al.}(2008){Nelson}, {Tomczyk}, {Elmore}, \&
  {Kolinski}}]{nelson:08}
{Nelson}, P.~G., {Tomczyk}, S., {Elmore}, D.~F., \& {Kolinski}, D.~J. 2008, in
  Society of Photo-Optical Instrumentation Engineers (SPIE) Conference Series,
  Vol. 7012, Society of Photo-Optical Instrumentation Engineers (SPIE)
  Conference Series

\bibitem[{{Roddier} \& {Roddier}(1993)}]{roddier:93}
{Roddier}, C. \& {Roddier}, F. 1993, Journal of the Optical Society of America
  A, 10, 2277

\bibitem[{{Slater} {et~al.}(2009){Slater}, {Harding}, \& {Mihos}}]{slater:09}
{Slater}, C.~T., {Harding}, P., \& {Mihos}, J.~C. 2009, \pasp, 121, 1267

\bibitem[{{Tal} {et~al.}(2009){Tal}, {van Dokkum}, {Nelan}, \&
  {Bezanson}}]{tal:09}
{Tal}, T., {van Dokkum}, P.~G., {Nelan}, J., \& {Bezanson}, R. 2009, \aj, 138,
  1417

\bibitem[{{Toomre} \& {Toomre}(1972)}]{toomre:72}
{Toomre}, A. \& {Toomre}, J. 1972, \apj, 178, 623

\bibitem[{{van Dokkum}(2005)}]{dokkum:05}
{van Dokkum}, P.~G. 2005, \aj, 130, 2647

\bibitem[{{Wallace}(1994)}]{wallace:94}
{Wallace}, P.~T. 1994, in Astronomical Society of the Pacific Conference
  Series, Vol.~61, Astronomical Data Analysis Software and Systems III, ed.
  D.~R. {Crabtree}, R.~J. {Hanisch}, \& J.~{Barnes}, 481

\end{thebibliography}

\end{document}